\documentclass[10pt]{article}
\usepackage{amssymb,amsmath,amsthm}
\usepackage{epsfig}
\usepackage[dvips]{color}
\definecolor{LightGreen1}{rgb}{0.2,0.8,0.2}
\definecolor{DarkGray}{rgb}{0.4,0.4,0.4}
\definecolor{GhostWhite}{rgb}{0.97,0.97,1.}
\newcommand{\ie}{\emph{i.e.}}
\newcommand{\eg}{\emph{e.g.}}
\newcommand{\cf}{\emph{cf}}
\newcommand{\etc}{\emph{etc}}
\newcommand{\viz}{\emph{viz}}
\newcommand{\Com}{\mathbb{C}}
\newcommand{\Real}{\mathbb{R}}
\newcommand{\Nat}{\mathbb{N}}
\newcommand{\Int}{\mathbb{Z}}
\newcommand{\sii}{L^2}

\newcommand{\Sobii}{\mathop{W^{2,2}}\nolimits}
\newcommand{\eps}{\varepsilon}
\newcommand{\PT}{\mathcal{PT}}
\newcommand{\iu}{\mathrm{i}}
\newtheorem{Theorem}{Theorem}
\newtheorem{Corollary}{Corollary}

\theoremstyle{definition}
\newtheorem*{Remark}{Remark}
\begin{document}
%
%
\title{\Large\textbf{%
Non-Hermitian spectral effects in a $\PT$-symmetric waveguide%
}}
\author{D.~Krej\v{c}i\v{r}\'{\i}k \ and \ M.~Tater}
\date{
\footnotesize
\begin{center}
\emph{
Department of Theoretical Physics, Nuclear Physics Institute,
\\
Academy of Sciences, 250\,68 \v{R}e\v{z} near Prague, Czech Republic
\smallskip \\
\emph{E-mails:} krejcirik@ujf.cas.cz, tater@ujf.cas.cz
}
\end{center}
7 February 2008}
\maketitle
\begin{abstract}
\noindent
We present a numerical study of the spectrum of the Laplacian
in an unbounded strip with $\PT$-symmetric boundary conditions.
We focus on non-Hermitian features of the model
reflected in an unusual dependence of the eigenvalues
below the continuous spectrum on various
boundary-coupling parameters.
\end{abstract}
%
%
%

\section{Introduction}
%
In the last years the theory of
quasi-Hermitian, pseudo-Hermitian and $\PT$-symmetric operators
has developed rapidly, and has been shown to provide
a huge class of non-Hermitian Hamiltonians with real spectra
(\cf~the pioneering works \cite{GHS, Bender-Boettcher_1998,Ali1}
and the review~\cite{Bender_2007} with many references).
Because of these recent observations,
the condition of self-adjointness of operators
representing observables in quantum mechanics
may seem to be rather an annoying technicality.
However, unless it is met one cannot apply the very powerful
machinery of spectral theory based on the spectral theorem.

In particular, one has to restrict to exactly solvable models
\cite{Levai-Znojil_2000, Znojil_2001, Swanson_2004, KBZ, Levai_2007}
or to rely on perturbation and numerical methods to analyse the spectrum
of the $\PT$-symmetric Hamiltonians.
Except for some simple one-dimensional examples~\cite{AFK},
the majority of the $\PT$-symmetric models
studied in the literature have purely discrete spectrum.
Perturbation methods are then notably effective in determining
the dependence of the eigenvalues on various parameters of the given model.
Although the perturbation approach can even prove
that the total spectrum is real in some cases
\cite{Langer-Tretter_2004,
CGS, Caliceti-Cannata-Graffi_2006, Caliceti-Graffi-Sjostrand_2007},
it is limited in its nature and one usually has to employ numerical techniques
in order to obtain a more complete picture of the spectral properties.

In a way motivated by the lack of a well-developed spectral theory
for non-self-adjoint operators with non-compact resolvent,
in a recent paper~\cite{BK} Borisov and one of the present
authors introduced a new two-dimensional $\PT$-symmetric Hamiltonian
with a real continuous spectrum.
One of the main questions arising within this model
is whether the Hamiltonian possesses point spectrum, too.
Using some singular perturbation techniques adopted from the theory of
quantum waveguides \cite{Ga,Bo}, the question was given both
positive and negative answer in~\cite{BK}, depending on the nature
of the effective $\PT$-symmetric interaction in a
\emph{weakly-coupled} regime. Moreover, in the case when the point
spectrum exists, the weakly-coupled eigenvalues emerging from the
continuous spectrum were shown to be real. We refer to the next
Section~\ref{Sec.model} for a precise statement of the spectral
results established in~\cite{BK}.

In the present paper, we further analyse the question of the
existence of point spectrum for the Hamiltonian introduced
in~\cite{BK} by \emph{numerical} methods.
This enables us to explore quantitative properties of the
eigenvalues without the restriction to the weakly-coupled regime.
The main emphasis is put on peculiar characteristics of the model
which are related to the non-self-adjointness of the underlying
Hamiltonian, namely:
\begin{enumerate}
\item
highly \emph{non-monotone} dependence of the eigenvalues on
a coupling parameter; as the parameter increases, the eigenvalues
emerge from the continuous spectrum, reach a minimum,
sometimes disappear in the continuous spectrum, emerge
later on again, \etc;
\item
\emph{broken $\PT$-symmetry};
as the coupling parameter increases,
the eigenvalues may emerge from the continuous spectrum
as complex-conju\-gate pairs, collide and become real,
move on the real axis,
collide again and become complex, \etc.
\end{enumerate}

This paper is organized as follows.
We begin by recalling the Hamiltonian from~\cite{BK}
and summarize the main spectral properties established there.
In particular, we point out some questions
the study of~\cite{BK} has left open.
In Section~\ref{Sec.numerics} we describe the numerical methods we use.
The numerical data are then presented
and discussed in Section~\ref{Sec.results}.
In the final Section~\ref{Sec.end}
we make some conjectures based on the present study.

\section{The model, known results and open questions}\label{Sec.model}
%
Given a positive number~$d$,
we introduce an infinite strip $\Omega:=\Real\times(0,d)$.
We split the variables consistently by writing
$x=(x_1,x_2)$ with $x_1\in\Real$ and $x_2\in (0,d)$.
Let $\alpha:\Real\to\Real$ be a bounded function;
occasionally we shall denote by the same symbol
the function $x\mapsto\alpha(x_1)$ on $\Omega$.
The Hamiltonian~$H_\alpha$ we consider in this paper
acts simply as the Laplacian in the Hilbert space $\sii(\Omega)$,
\ie
\begin{equation}
  H_\alpha\Psi := -\Delta\Psi
  \qquad\mbox{in}\quad
  \Omega
  \,,
\end{equation}
and a non-trivial interaction is introduced
by choosing as its domain the set of functions~$\Psi$
from $\Sobii(\Omega)$
satisfying the following Robin-type boundary conditions:
\begin{equation}\label{bc}
  \partial_2\Psi + \iu\,\alpha\,\Psi
  \, = \, 0
  \qquad\mbox{on}\quad
  \partial\Omega
  \,.
\end{equation}
Here $\Sobii(\Omega)$ denotes the Sobolev space
consisting of functions on~$\Omega$ which,
together with all their first and second distributional derivatives,
are square integrable.
As usual, the action of~$H_\alpha$ should be understood
in the distributional sense
and~\eqref{bc} should be understood
in the sense of traces~\cite{Adams}.

\begin{figure}[h]
\epsfig{file=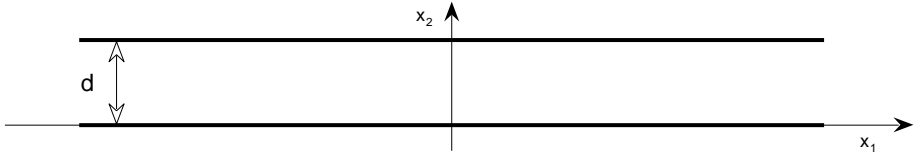, width=\textwidth} \caption{A schematic
view of an infinite planar waveguide of width $d$. The Robin
conditions~\eqref{bc} are imposed at the boundary.}
\end{figure}

Under the additional hypothesis
that~$\alpha$ possesses a bounded distributional derivative,
\ie~$\alpha \in W^{1,\infty}(\Real)$,
it was shown in~\cite{BK} that~$H_\alpha$ is an $m$-sectorial operator
satisfying
\begin{equation}\label{symmetry}
  H_\alpha^* = H_{-\alpha}
  \,,
\end{equation}
where~$H_\alpha^*$ denotes the adjoint of~$H_\alpha$.
(If~$\alpha$ is merely bounded, it is still possible
to give a meaning to~$H_\alpha$ by using the quadratic-form approach.)
Of course, $H_\alpha$~is not self-adjoint
unless~$\alpha$ vanishes identically.
However, the relation~\eqref{symmetry} reflects the $\PT$-symmetry
-- or, more generally and more precisely,
the $\mathcal{T}$-self-adjointness -- of~$H_\alpha$,
with~$\mathcal{P}$ and~$\mathcal{T}$
being defined by $(\mathcal{P}\Psi)(x):=\Psi(d-x)$
and $\mathcal{T}\Psi:=\overline{\Psi}$, respectively.

An important property of the operator~$H_\alpha$
being $m$-sectorial is that it is closed.
Then, in particular, the spectrum~$\sigma(H_\alpha)$
is well defined as the set of complex points~$z$
such that $H_\alpha-z$ is not bijective.
The point spectrum~$\sigma_\mathrm{p}(H_\alpha)$
equals the set of points~$z$ such that $H_\alpha-z$ is not injective.
The continuous spectrum~$\sigma_\mathrm{c}(H_\alpha)$
equals the set of points~$z$ such that $H_\alpha-z$ is not surjective
but the range of $H_\alpha-z$ is dense in $\sii(\Omega)$.
Finally, the residual spectrum~$\sigma_\mathrm{r}(H_\alpha)$
equals the set of points~$z$ such that $H_\alpha-z$ is injective
but the range of $H_\alpha-z$ is not dense in $\sii(\Omega)$.

In the following theorem we collect general results
about the spectrum of~$H_\alpha$ established in~\cite{BK}:
\begin{Theorem}\label{Thm1}
Let $\alpha\in W^{1,\infty}(\Real)$ and $\alpha_0\in\Real$.
Then
\begin{itemize}
\item[\emph{(i)}]
$
  \sigma(H_\alpha) \subseteq
  \{z\in\Com \,:\, |\arg(z)| \leq \theta\}
$
with some $\theta \in [0,\pi/2)$;
\item[\emph{(ii)}]
$
  \sigma_\mathrm{r}(H_\alpha)=\varnothing
$;
\item[\emph{(iii)}]
$
  \sigma(H_{\alpha_0})
  = \sigma_\mathrm{c}(H_{\alpha_0})
  = [\mu_0^2,\infty)
$
\ where \
$\mu_0:=\min\{|\alpha_0|,\pi/d\}$;
\item[\emph{(iv)}]
if $\alpha-\alpha_0 \in C_0(\Real)$, then \
$
  \sigma_\mathrm{c}(H_{\alpha})
  = [\mu_0^2,\infty)
$;
\item[\emph{(v)}]
if $\alpha \in C_0(\Real)$ is an odd function, then \
$
  \sigma_\mathrm{p}(H_{\alpha})
  \subset \Real
$;
\end{itemize}
\end{Theorem}
\noindent
Here $C_0(\Real)$ denotes the space of continuous functions
on~$\Real$ with compact support.
Note that $\alpha \in C_0(\Real) \cap W^{1,\infty}(\Real)$
implies that~$\alpha$ is Lipschitz continuous;
conversely, the space of Lipschitz continuous functions on~$\Real$
is embedded in $W^{1,\infty}(\Real)$.

The first two properties of Theorem~\ref{Thm1} are quite general:
(i)~holds since~$H_\alpha$ is sectorial
and (ii)~is a consequence of the~$\mathcal{T}$-self-adjointness
(we refer to~\cite{BK} for more details).
Since the spectral problem for the constant case of~(iii)
can be solved by some sort of ``separation of variables''
(\cf~\cite[Sec.~4]{BK}),
we shall refer to it as the unperturbed case;
it follows from Theorem~\ref{Thm1}
that the corresponding spectrum is purely
continuous and positive.
As a consequence of (ii), (iv) and~(v),
we get a result about the reality of the total spectrum
in the perturbed case:
\begin{Corollary}\label{Corol.reality}
Let $\alpha\in C_0(\Real) \cap W^{1,\infty}(\Real)$ be an odd function.
Then
$$
  \sigma(H_{\alpha})
  \subset \Real
  \,.
$$
\end{Corollary}

The result stated in part~(iv) of Theorem~\ref{Thm1}
makes rigorous the heuristic statement that
``the continuous spectrum depends on the properties
of a Hamiltonian interaction at infinity only''.
On the other hand, it is well known
-- and this already for one-dimensional self-adjoint models~\cite{Kl} --
that the point spectrum may be highly unstable under a perturbation
of an operator with non-compact resolvent.
In~\cite{BK}, the point spectrum of~$H_\alpha$
was analysed perturbatively in the weakly-coupled regime:
\begin{equation}\label{weak.coupling}
  \alpha = \alpha_0+\eps\,\beta \,,
  \qquad\mbox{with}\quad
  \alpha_0 \in \Real \,, \
  \eps > 0 \,, \
  \beta\in C_0^2(\Real) \,, \
\end{equation}
where~$\beta$ is assumed to be real-valued
and~$\eps$ plays the role of the small parameter.
Here $C_0^2(\Real)$ denotes the space of functions on~$\Real$ which,
together with all their first and second derivatives,
are continuous and have compact support.
The main interest was focused on the existence
and asymptotic behavior of the eigenvalues emerging from
the threshold~$\mu_0^2$ of the continuous spectrum
due to the perturbation of~$H_{\alpha_0}$ by~$\eps\beta$.

Before stating the main results of~\cite{BK} about the
weakly-coupled eigenvalues, we need to introduce
some notation.
Recalling the definition of~$\mu_0$ from Theorem~\ref{Thm1}.(iii),
we next define
$$
  \mu_1:=\max\{|\alpha_0|,\pi/d\}
  \qquad\mbox{and}\qquad
  \mu_j:=\pi j/d
  \quad\mbox{for} \quad j\geq 2 \,.
$$
To these numbers we associate a family of functions
$\{\psi_j\}_{j=0}^\infty$ by
\begin{equation}\label{psi}
  \psi_j(x_2) :=
  \cos(\mu_j x_2) - \iu \, \frac{\alpha_0}{\mu_j}\, \sin(\mu_j x_2)
  \,.
\end{equation}
Let $\{v_j\}_{j=0}^\infty$ be a sequence of auxiliary functions given by
\begin{equation*}
  v_j(x_1) :=\left\{
  \begin{aligned}
    &-\frac{1}{2}\int_\Real |x_1-t_1| \beta(t_1) \, d t_1
    & \text{if} \quad j=0 \,,
    \\
    &\frac{1}{2\sqrt{\mu_j^2-\mu_0^2}}
    \int_\Real e^{-\sqrt{\mu_j^2-\mu_0^2}|x_1-t_1|}
    \beta(t_1) \, d t_1
    & \text{if} \quad j \geq 1 \,.
  \end{aligned}\right.
\end{equation*}
Finally, denoting $\langle f\rangle=\int_\Real f(x_1) \, d x_1$
for any $f \in L^1(\Real)$, we introduce a constant~$\tau$,
depending on~$\beta$, $d$ and~$\alpha_0$, by
\begin{equation}\label{tau}
  \tau :=\left\{
  \begin{aligned}
  &2\alpha_0^2\langle\beta v_0\rangle
  +\frac{2\alpha_0}{d}
  \sum\limits_{j=1}^{\infty}
  \frac{\mu_j^2\langle\beta v_j\rangle}{\mu_j^2-\mu_0^2}
  \, \tan\frac{\alpha_0 d+j\pi}{2} \
  & \text{if} \quad
  |\alpha_0| < \frac{\pi}{d}
  \,,
  \\
  &\frac{2\alpha_0\pi^2\cot\frac{\alpha_0 d}{2}}{(\mu_1^2-\mu_0^2)d^3}
  \langle\beta v_1\rangle + \frac{8 \pi^2}{(\mu_1^2-\mu_0^2)d^4}
  \sum\limits_{j=1}^{\infty} \frac{\mu_{2j}^2\langle \beta v_{2j}
  \rangle}{\mu_{2j}^2-\mu_1^2} \
  & \text{if} \quad
  |\alpha_0| > \frac{\pi}{d}
  \,.
  \end{aligned}\right.
\end{equation}
Now we are in a position to take over from~\cite{BK}:
\begin{Theorem}\label{Thm2}
Let~$\alpha$ be given by~\eqref{weak.coupling}.
\begin{itemize}
\item[\emph{(A)}]
If $\alpha_0 = 0$, then $H_\alpha$ has no eigenvalues
converging to $\mu_0^2$ as $\eps\to 0$.
\item[\emph{(B)}]
Let $0<|\alpha_0|<\pi/d$.
\begin{enumerate}
\item[\emph{1.}]
If $\alpha_0\langle\beta\rangle<0$, then there exists the unique eigenvalue
$\lambda_\eps$ of $H_\alpha$ converging to $\mu_0^2$ as $\eps \to 0$.
This eigenvalue is simple and real, and satisfies the asymptotic
formula
\begin{equation}\label{asB}
\lambda_\eps=\mu_0^2-\eps^2\alpha_0^2\langle\beta\rangle^2
+2\eps^3\alpha_0\tau\langle \beta \rangle
+\mathcal{O}(\eps^4) \,.
\end{equation}
\item[\emph{2.}]
If $\alpha_0\langle\beta\rangle>0$,
then $H_\alpha$ has no eigenvalues
converging to $\mu_0^2$ as $\eps \to 0$.
\item[\emph{3.}]
If $\langle\beta\rangle=0$ and $\tau>0$, then there exists the unique
eigenvalue $\lambda_\eps$ of $H_\alpha$ converging to $\mu_0^2$ as
$\eps\to 0$. This eigenvalue is simple and real, and satisfies the
asymptotics
\begin{equation}\label{asB.critical}
\lambda_\eps=\mu_0^2-\eps^4\tau^2+\mathcal{O}(\eps^5) \,.
\end{equation}
\item[\emph{4.}]
If $\langle\beta\rangle=0$ and $\tau<0$, then $H_\alpha$ has no
eigenvalues converging to~$\mu_0^2$ as $\eps\to 0$.
\end{enumerate}
\item[\emph{(C)}]
Let $|\alpha_0|>\pi/d$ \ and \ $\alpha_0 d/\pi \not\in \Int$.
\begin{enumerate}
\item[\emph{1.}]
If $\tau>0$, then there exists the unique eigenvalue $\lambda_\eps$
of $H_\alpha$ converging to $\mu_0^2$ as $\eps\to 0$, it is simple and
real, and satisfies the asymptotics~\eqref{asB.critical}.
\item[\emph{2.}]
If $\tau<0$, then $H_\alpha$ has no eigenvalues converging
to $\mu_0^2$ as $\eps \to 0$.
\end{enumerate}
\end{itemize}
\end{Theorem}

The method of~\cite{BK} gives also the asymptotic
expansion of the eigenfunctions corresponding to the
weakly-coupled eigenvalues:
\begin{Theorem}\label{Thm3}
The eigenfunction~$\Psi_\eps$ corresponding to
any eigenvalue~$\lambda_\eps$ from Theorem~\ref{Thm2}
can be chosen so that it satisfies the asymptotics
\begin{equation}\label{1.3}
  \Psi_\eps(x)=\psi_0(x_2)+\mathcal{O}(\eps)
\end{equation}
in $\Sobii(\Omega\cap\{x: |x_1|<a\})$ for each $a>0$,
and behaves at infinity as
\begin{equation}\label{1.4}
\Psi_\eps(x)=\exp^{-\sqrt{\mu_0^2-\lambda_\eps}|x_1|}\psi_0(x_2)+
\mathcal{O}(\exp^{-\sqrt{\mu_0^2-\lambda_\eps}|x_1|}) \,,
\quad x_1\to+\infty \,.
\end{equation}
\end{Theorem}

Theorems~\ref{Thm1}--\ref{Thm3} summarizing the spectral analysis
performed in~\cite{BK} leave open the following particular questions:
\begin{enumerate}
\item[(Q1)]
Can the cases (B4) and (C2) of Theorem~\ref{Thm2} occur?
That is, can the constant~$\tau$ be negative for a certain
combination of $d, \alpha_0$ and~$\beta$?
(Sufficient conditions for the positivity of~$\tau$ exist
\cite[Props.~2.1--2.2]{BK}.)
\item[(Q2)]
What happens in the case (C) of Theorem~\ref{Thm2}
if the condition $\alpha_0 d/\pi \not\in \Int$ is not satisfied?
Is it just a technical hypothesis?
\item[(Q3)]
Is there any point spectrum in the case of Corollary~\ref{Corol.reality}?
\item[(Q4)]
What is the dependence of the weakly-coupled eigenvalues
of Theorem~\ref{Thm2} as the parameter~$\eps$ increases?
\item[(Q5)]
Do the eigenvalues remain real for large~$\eps$?
\item[(Q6)]
Can one have more eigenvalues? Can they be degenerate?
What is the dependence of the number of eigenvalues on~$\eps$?
\item[(Q7)]
Are there any eigenvalues emerging
from other thresholds $\mu_j^2$, $j \geq 1$?
Can they emerge from other points of the continuous spectrum,
different from the thresholds $\mu_j^2$, $j \geq 0$?
\end{enumerate}
The main goal of the present paper is to provide answers to
some of these questions by a numerical study of the spectral problem.

\section{Numerical methods}\label{Sec.numerics}
%
In order to get the dependence of the bound states on parameters
like $\eps, d, \alpha_0$, \etc,
numerically we used two independent methods.
When $\alpha$ is a simple step-like function
(\eg\ symmetric or asymmetric square well),
we treat the problem by mode matching method.
It takes into account the asymptotic
behaviour of solution explicitly and can serve thus as a useful
check when we apply the other method, \viz, the spectral method.
This method is more robust and we use it for more general
$\alpha$. We arrived at an excellent agreement in cases when both
methods are applicable.

\subsection{Mode matching method}\label{Sec.matching}
\begin{figure}[h!]
\epsfig{file=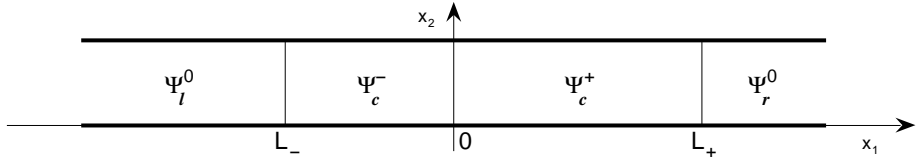, width=\textwidth} \caption{The mode
matching approach. A particular \emph{Ansatz}~\eqref{Ansatz} for
an eigenfunction~$\Psi$ of~$H_\alpha$ corresponding to~$\lambda$
is chosen in each subregion and the smooth matching~\eqref{match}
is required at the boundaries separating the
subregions.}\label{Fig.mode}
\end{figure}

%
Let us begin with mode matching. The most general situation we
want to describe is shown in Figure~\ref{Fig.mode}.
Fix negative and positive numbers~$L_-$ and~$L_+$, respectively.
In the asymptotic regions,
\ie\ $x_1<L_-$ and $L_+<x_1$,
we assume $\alpha(x_1)=\alpha_0$,
while in the central parts
we have $\alpha(x_1)=\alpha_-$ if $L_-<x_1<0$
and $\alpha(x_1)=\alpha_+$ if $0<x_1<L_+$.

Let $\{\mu_j^\pm\}_{j=0}^\infty$ denote the sequence of numbers
$\{\mu_j\}_{j=0}^\infty$ with $\alpha_0$ being replaced by~$\alpha_\pm$.
In the same way we define the sequence of functions
$\{\psi_j^\pm\}_{j=0}^\infty$ by replacing~$\alpha_0$ by~$\alpha_\pm$
in~\eqref{psi}.
In order to make the notation more consistent,
hereafter we write $\mu_j^0$ and $\psi_j^0$ instead of
$\mu_j$ and $\psi_j$, respectively,
and introduce a common index $\iota\in\{0,+,-\}$.

In each of the regions where~$\alpha$ is constant,
the spectral problem $-\Delta\Psi = \lambda\Psi$,
with~$\Psi$ satisfying the required boundary conditions,
can be solved explicitly \cite[Sec.~4]{BK}
by expanding~$\Psi$
into the ``transverse basis'' $\{\psi_j^\iota\}_{j=0}^\infty$,
where~$\iota$ depends on the region.
More specifically, we use the following \emph{Ansatz}
for an eigenfunction~$\Psi$ of~$H_\alpha$
corresponding to~$\lambda$:
\begin{equation}\label{Ansatz}
  \Psi(x) =
  \left\{
  \begin{aligned}
    \Psi_l^0(x) &:=
    \sum_{j=0}^\infty d_j \,
    e^{\sqrt{(\mu_j^{0})^2-\lambda}\,x_1} \,
    \psi_j^0(x_2)
    & \mbox{if} & \quad x_1 \in (-\infty,L_-) \,, \\
    \Psi_c^-(x) &:=
    \sum_{j=0}^\infty c_j \, \varphi_j^-(x_1) \, \psi_j^-(x_2)
    & \mbox{if} & \quad x_1 \in (L_-,0) \,, \\
    \Psi_c^+(x) &:=
    \sum_{j=0}^\infty b_j \, \varphi_j^+(x_1) \, \psi_j^+(x_2)
    & \mbox{if} & \quad x_1 \in (0,L_+) \,, \\
    \Psi_r^0(x) &:=
    \sum_{j=0}^\infty a_j \,
    e^{-{\sqrt{(\mu_j^{0})^2-\lambda}\,x_1}} \,
    \psi_j^0(x_2)
    & \mbox{if} & \quad x_1 \in (L_+,+\infty) \,, \\
  \end{aligned}
  \right.
\end{equation}
where
\begin{equation*}
  \varphi_j^\pm(x_1) :=
  \cos\big(\sqrt{\lambda-(\mu_j^\pm)^2}\,x_1\big)
  +B_\pm \, \sin\big(\sqrt{\lambda-(\mu_j^\pm)^2}\,x_1\big)
  \,.
\end{equation*}

Standard elliptic regularity theory implies that
any weak solution~$\Psi$ to $-\Delta\Psi=\lambda\Psi$
is necessarily infinitely smooth in the interior of~$\Omega$.
In particular, we must match the functions from the \emph{Ansatz}
smoothly at $x_1=L_-,0,L_+$, \ie\ we require
\begin{align}\label{match}
  \Psi_l^0(L_-,x_2)&=\Psi_c^-(L_-,x_2) \,,
  &
  \partial_1\Psi_l^0(L_-,x_2)&=\partial_1\Psi_c^-(L_-,x_2) \,,
  \nonumber \\
  \Psi_c^-(0,x_2)&=\Psi_c^+(0,x_2) \,,
  &
  \partial_1\Psi_{c}^-(0,x_2)&=\partial_1\Psi_{c}^+(0,x_2) \,,
  \\
  \Psi_c^+(L_+,x_2)&=\Psi_r^0(L_+,x_2) \,,
  &
  \partial_1\Psi_{c}^+(L_+,x_2)&=\partial_1\Psi_{+}^0(L_+,x_2) \,,
  \nonumber
\end{align}
for every $x_2\in(0,d)$.

If $\{\psi_j^\iota\}_{j=0}^\infty$ formed an orthonormal family,
the next step would consist in employing the orthonormality
and reducing~\eqref{match} into a system of algebraic equations
for the coefficients $a_j,b_j,c_j,d_j$.
However, since the family $\{\psi_j^\iota\}_{j=0}^\infty$
is actually formed by eigenfunctions of a transverse eigenvalue problem
which is not Hermitian (unless $\alpha_\iota=0$),
it is clear that the functions~$\psi_j^\iota$
are not mutually orthogonal in general.
Instead, we use the property that $\{\psi_j^\iota\}_{j=0}^\infty$
and $\{\phi_j^\iota\}_{j=0}^\infty$
form a complete biorthonormal pair \cite{KBZ},
where $\phi_j^\iota$ are properly normalized eigenfunctions
of the adjoint transverse problem:
\begin{equation*}
  \phi_j^\iota(x_2) := \overline{A_j^\iota \, \psi_j^\iota(x_2)}
  \,.
\end{equation*}
The normalization constants can be chosen as follows
\begin{equation*}
  A_{j_0}^\iota
  :=\frac{2\iu\alpha_\iota}{1-\exp{(-2\iu\alpha_\iota d)}}
  \,, \quad
  A_{j_1}^\iota
  :=\frac{2 (\mu_1^\iota)^2}{[(\mu_1^\iota)^2-\alpha_\iota^2]d}
  \,, \quad
  A_j^\iota
  :=\frac{2(\mu_j^\iota)^2}{[(\mu_j^\iota)^2-\alpha_\iota^2]d}
  \,,
\end{equation*}
where $j \geqslant 2$,
$(j_0,j_1) = (0,1)$ if $|\alpha_\iota|<\pi/d$
and $(j_0,j_1) = (1,0)$ if $|\alpha_\iota|>\pi/d$
(if $\alpha_\iota=0$, the fraction in the definition of $A_{j_0}^\iota$
should be understood as the expression obtained after taking the
limit $\alpha_\iota \to 0$).
Then, in particular, we have
\begin{equation}\label{orthonormality}
  \forall i,j\in\Nat, \quad
  (\phi_i^\iota,\psi_j^\iota) = \delta_{ij}
  \,,
\end{equation}
where $(\cdot,\cdot)$ denotes the inner product in $\sii((0,d))$,
antilinear in the first factor and linear in the second one.

Now, multiplying~\eqref{match} by~$\overline{\phi_i^0}$,
integrating over $x_2\in(0,d)$ and employing~\eqref{orthonormality}
in the asymptotic regions, we can eliminate
the coefficients $a_j$ and $d_j$ by means of the relations
\begin{equation}\label{ajdj}
\begin{aligned}
  a_i \,
    e^{-\sqrt{(\mu_i^{0})^2-\lambda}\,L_+}
    &=\sum_{j=0}^\infty b_j \,
    \varphi_j^+(L_+) \, (\phi_i^0,\psi_j^+) \,,
  \\
  d_i \,
    e^{\sqrt{(\mu_i^{0})^2-\lambda}\,L_-}
    &=\sum_{j=0}^\infty c_j \,
    \varphi_j^-(L_-) \, (\phi_i^0,\psi_j^-) \,,
\end{aligned}
\end{equation}
for every $i \in \Nat$,
and reduce thus the number of conditions to be fulfilled.
We finally arrive at an infinite-dimensional homogeneous system
\begin{equation}\label{system}
\left(
\begin{array}{c c c c}
m_{11} & m_{12} & 0 & 0 \\
m_{21} & 0 & m_{23} & 0 \\
0 & 0 & m_{33} & m_{34} \\
0 & m_{42} & 0 & m_{44}
\end{array}
\right) \left(
\begin{array}{c}
b \\
c \\
bB_+ \\
cB_-
\end{array} \right)
= \left(
\begin{array}{c}
0 \\
0 \\
0 \\
0
\end{array} \right)\,.
\end{equation}
Here $b,c$ denote the infinite vectors formed by $b_j,c_j$,
respectively, and the submatrices~$m_{\mu\nu}$ are given by
\begin{align*}
  m_{11}&:=(\phi_i^0,\psi_j^+) \,,
  &
  m_{33}&:=\sqrt{\lambda-(\mu_j^+)^2} \, (\phi_i^0,\psi_j^+) \,,
  \\
  m_{12}&:=-(\phi_i^0,\psi_j^-) \,,
  &
  m_{34}&:=-\sqrt{\lambda-(\mu_j^-)^2} \, (\phi_i^0,\psi_j^-) \,,
\end{align*}
\begin{align*}
  m_{21}
  &:= \bigg(
  \mbox{$\sqrt{(\mu_i^0)^2-\lambda}$} \,
  \cos\big(L_+\sqrt{\lambda-(\mu_j^+)^2}\big)
  \\
  & \qquad\qquad
  -\sqrt{\lambda-(\mu_j^+)^2} \,
  \sin\big(L_+\sqrt{\lambda-(\mu_j^+)^2}\big)
  \bigg)
  (\phi_i^0,\psi_j^+) \,,
  \\
  m_{23}
  &:= \bigg(
  \sqrt{(\mu_i^0)^2-\lambda} \,
  \sin\big(L_+\sqrt{\lambda-(\mu_j^+)^2}\big)
  \\
  & \qquad\qquad
  +\sqrt{\lambda-(\mu_j^+)^2} \,
  \cos\big(L_+\sqrt{\lambda-(\mu_j^+)^2}\big)
  \bigg)
  (\phi_i^0,\psi_j^+) \,,
  \\
  m_{42}
  &:= \bigg(
  \sqrt{(\mu_i^0)^2-\lambda} \,
  \cos\big(L_-\sqrt{\lambda-(\mu_j^-)^2}\big)
  \\
  & \qquad\qquad
  +\sqrt{\lambda-(\mu_j^-)^2} \,
  \sin\big(L_-\sqrt{\lambda-(\mu_j^-)^2}\big)
  \bigg)
  (\phi_i^0,\psi_j^-) \,,
  \\
  m_{44}
  &:= \bigg(
  \sqrt{(\mu_i^0)^2-\lambda} \,
  \sin\big(L_-\sqrt{\lambda-(\mu_j^-)^2}\big)
  \\
  & \qquad\qquad
  -\sqrt{\lambda-(\mu_j^-)^2} \,
  \cos\big(L_-\sqrt{\lambda-(\mu_j^-)^2}\big)
  \bigg)
  (\phi_i^0,\psi_j^-) \,,
\end{align*}
where the right hand sides should be understood
as the infinite matrices formed by the respective coefficients
for $i,j\in\Nat$.
Our numerical approximation then consists in approximating
the infinite system by using finite submatrices
for $i,j\in\{0,\dots,N\}$ with $N$ large enough.

In order to have a nontrivial solution we require
\begin{equation}\label{implicit}
\det
\left(
\begin{array}{c c c c}
m_{11} & m_{12} & 0 & 0 \\
m_{21} & 0 & m_{23} & 0 \\
0 & 0 & m_{33} & m_{34} \\
0 & m_{42} & 0 & m_{44}
\end{array}
\right)
= 0
\,,
\end{equation}
which gives an implicit equation for~$\lambda$ as the unknown.
Having found~$\lambda$, we can then calculate the coefficients
$a_j,b_j,c_j,d_j,B_{\pm,j}$ from~\eqref{system} and~\eqref{ajdj}.

If $\alpha_+=\alpha_-$ and $L_+=-L_-$,
\ie \ $\alpha$ is a symmetric square well,
then~\eqref{implicit} can be reduced to
$$
\det\left(
\begin{array}{c c}
m_{21}+m_{42} & 0 \\
0 & m_{23}-m_{44}
\end{array}
\right)=0
\,.
$$
The solutions have different symmetry with respect to $x_1 \mapsto
-x_1$, the even solutions are formed only of cosines, the odd of
sines.

\begin{Remark}
In principle, it is possible to extend the present method to an
arbitrary piece-wise constant function~$\alpha$, provided that the
number of matching interfaces is finite. On the other hand, the
more matching conditions the bigger size of the matrix
of~\eqref{system} and thus the higher (numerical) price one must
pay.
\end{Remark}
%

\subsection{Spectral method}
%
In order to treat the waveguide with a general $\alpha$ in the
boundary conditions, we decided to use spectral collocation
methods. They provide a reliable and rapidly converging tool
easily applicable to our model. A very useful software suite has
already been published~\cite{WR}. It could be adapted to this
problem. We approximate the operator by a series of operators
defined on a finite domain $[x_1^{min},x_1^{max}]\times[0,d]$ with
Dirichlet boundary conditions on $\{x_1^{min}\}\times[0,d]$ and
$\{x_1^{max}\}\times[0,d]$.

First, we can form the differentiation matrices in each variable
separately and then combine them to the two-dimensional problem.
The infinite domain in $x_1$-variable suggests that
we could approximate it by grid points
chosen as the roots of Hermite polynomials
and as an interpolant we take Lagrange polynomial. There is an
additional parameter (the real line can be mapped to itself by a
change of variable $x_1=b\tilde{x}_1, b>0$), which can be used to
optimize the choice of the grid points together with variation of
the number of roots $N_1$.
The roots span the interval
$[\xi_1,\xi_{N_1}]$, $-\xi_1=\xi_{N_1}$, cluster around the
origin, and grow as $\xi_{N_1}=\mathcal{O}(\sqrt{N_1})$ for
$N_1\rightarrow\infty$.

Another possibility is to use Fourier differencing. We form a
uniform grid in $[-x_1^{max},x_1^{max}]$ and since the solutions
decay exponentially, we can theoretically extend it periodically
across this interval to the whole $\mathbb{R}$. The interpolant is
a trigonometric function.

In both approaches the interpolant is an infinitely differentiable
function. Deriving it and taking the derivatives in the grid
points we get the differentiation matrices. Imposing the
homogeneous Dirichlet boundary conditions consists in deleting the
first and the last rows and columns of the differentiation
matrices.

The transversal variable is confined to a finite interval $[0,d]$
and it is possible to scale it to $[-1,1]$. To implement the
boundary conditions we prefer to incorporate them into the
interpolant. It requires to use Hermite interpolation, which takes
into account derivative values in addition to function values.
The use of the roots of Chebyshev polynomials
$\eta_k=\cos((k-1)\pi/(N_2-1)),
k=1,\ldots,N_2$ as the grid points is common here. For details we
refer the reader to~\cite{WR}.

Now, it remains to form differentiation matrices that correspond
to partial derivatives entering the Laplacian. Having set up a
grid in each direction we combine them into the tensor product grid.
Then a closer inspection shows that $
  \partial_{1}^2
  \rightarrow
  D^{(2)}(x_1)\otimes \mathbf{I}
$,
where $\mathbf{I}$ is an $N_2\times N_2$ identity matrix
and $\partial_{2}^2$ is constructed in a similar way
(it is necessary to take into
account that the boundary conditions change with~$x_1$).

Applying the spectral discretization we converted the search for
eigenvalues of $H_\alpha$ to a matrix eigenvalue problem.

\section{Discussion of numerical results}\label{Sec.results}
%
Existence of eigenvalues below the threshold of the continuous
spectrum and their behaviour for weak perturbations was already
proved in~\cite{BK}. Our calculations confirm it and demonstrate
that the spectrum of eigenvalues is considerably richer.

\begin{figure}[h!]
\epsfig{file=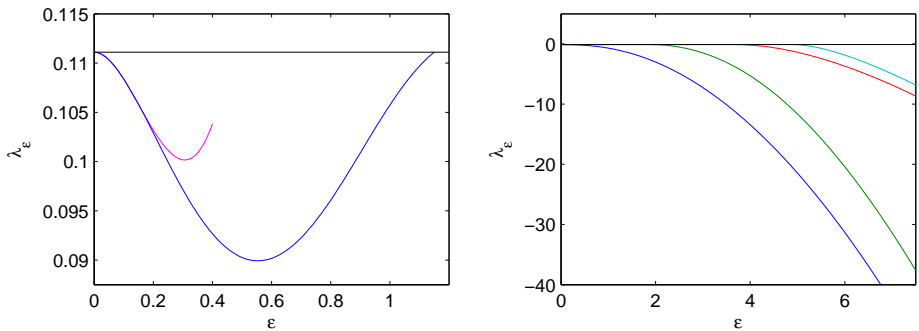, width=\textwidth}
\caption{Comparison of dependence of eigenvalues
on~$\eps$ for $\PT$-symmetric and self-adjoint waveguides.
The left figure shows the $\PT$-symmetric case
with $\alpha(x_1)=1/3-\eps \exp(-x_1^2)$.
Here the blue (respectively magenta) curve represents
the eigenvalue (respectively the asymptotic formula~\eqref{asB}
up to the $\eps^3$-term).
The right figure shows the eigencurves
in the corresponding self-adjoint situation,
obtained by replacing $\iu\alpha\mapsto \alpha$ in~\eqref{bc}.
$d=2$ in both cases.}
 \label{Fig.typical}
\end{figure}

A typical dependence of an eigenvalue on the perturbation
parameter~$\eps$ is shown in Figure~\ref{Fig.typical}.
Here we perturbed a waveguide of width $d=2$
and $\alpha_0=1/3$ by a Gaussian shape,
\ie\ we took $\beta(x_1)=-\exp(-x_1^2)$ in~\eqref{weak.coupling}.
Since $0<\alpha_0<\pi/d$ and $\langle\beta\rangle<0$,
we deal with the case~(B1) of Theorem~\ref{Thm2}.
We observe that the asymptotic expansion~\eqref{asB} is fairly good.
It is striking, however, that the dependence of the eigenvalue
on~$\eps$ is highly non-monotonic:
The eigenvalue appears at some value of $\eps$
(in this case it is $\eps=0$), reaches a minimum,
and then returns to the continuous spectrum.
We found such a behaviour in all cases we studied,
\viz, various shapes of symmetric and asymmetric wells,
and Gaussians times polynomials.
This provides an interesting answer to~(Q4)
from the end of Section~\ref{Sec.model}.

It is worth noting that this behaviour differs from that
in the self-adjoint waveguide obtained simply
by omitting the imaginary unit in~\eqref{bc}.
As shows the second graph in Figure~\ref{Fig.typical},
in the self-adjoint case all the energy levels
are increasingly more bound when~$\eps$ increases.

On the other hand, we checked that
the eigenvalues are decreasing as functions of $L:=\pm L_\pm$
for the symmetric square-well profile
$\alpha_+=\alpha_-$ of Section~\ref{Sec.matching}
in the regime $0<\alpha_\pm<\alpha_0<\pi/d$.
This is reasonable to expect since as $L\to\infty$
the eigenvalues should approach~$(\alpha_\pm)^2$,
\ie\ the threshold of the continuous spectrum
of the unperturbed waveguide~$H_{\alpha_\pm}$.

\begin{figure}[h!]
\epsfig{file=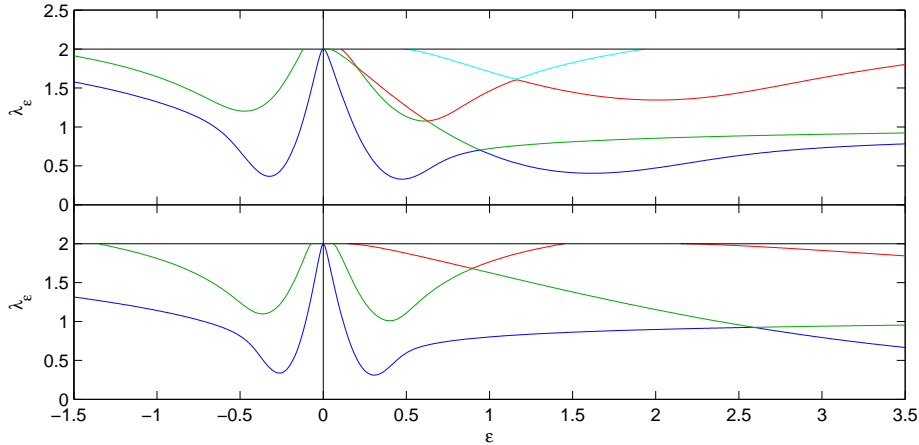, width=\textwidth}
\caption{Dependence of eigenvalues on~$\eps$
in the critical case $\langle\beta\rangle=0$, $\tau>0$.
Here $\alpha(x_1)=\sqrt{2}-\eps (x_1^2 + b x_1 - 5) \exp(-x_1^2/10)$
and $d=2$.
The upper figure corresponds to $b=1.5$, the lower one to $b=3.25$.
All the crossings are avoided.}\label{Fig.crossing}
\end{figure}

An answer to questions from~(Q6) is provided by Figure~\ref{Fig.crossing}.
It corresponds to the critical case (B3) of Theorem~\ref{Thm2} with
$\beta(x_1)=-(x_1^2+bx_1-5)\exp(-x_1^2/10)$,
$\alpha_0=\sqrt{2}$,
and $d=2$;
the parameter~$b$ changes the asymmetry of~$\beta$.
In addition to the weakly-coupled eigenvalue of Theorem~\ref{Thm2},
there are also other eigenvalues emerging from the continuous spectrum
as~$\eps$ increases.
The lower figure (case $b=3.25$) shows that there might be eigenvalues
existing in disjunct intervals of~$\eps$ (the red curve).
By diminishing~$\alpha_0$ we can achieve the situation when there is
only one eigenvalue with a similar behaviour:
it emerges from the threshold of continuous spectrum (at $\eps=0$),
reaches a minimum, returns to the continuum,
reappears later on, and returns finally to the continuum.

Another typical feature is that the energy levels do not cross.
We saw always avoided crossings
(at least in the unbroken $\PT$-regime),
\ie\ the order of levels remains
unchanged and the non-monotonicity of the excited eigenvalues is preserved.

\begin{figure}[h!]
\epsfig{file=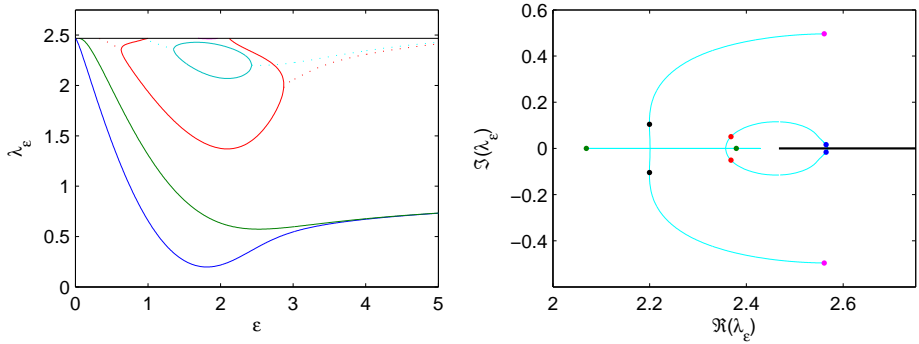, width=\textwidth}
\caption{Broken $\PT$-symmetry.
The left figure shows the dependence of
eigenvalues on~$\eps$ in the case $\alpha(x_1)=\pi/2-\eps
\exp(-x_1^2/10)$ and $d=2$. Here the dotted line is used to plot
the real part of the eigenvalues if they form complex conjugate
pairs instead of being real. The right figure shows the trajectory
of a pair of (complex) eigenvalues in the complex plane.
Here the thick black line marks the continuous spectrum.
The pairs of dots show positions of eigenvalues
for different values of~$\eps$:
$0.1$ (blue),
$1.3$ (red),
$2.1$ (green),
$2.5$ (black),
and $8$ (magenta).
An animation can be found at the website~\cite{video}.}\label{Fig.broken}
\end{figure}

The spectrum in Figure~\ref{Fig.broken} is
remarkable from two points of view.
First, it corresponds to the case of~(Q2)
from the end of Section~\ref{Sec.model},
since the constant $\alpha_0$ is chosen so that its square
coincides with the threshold of the continuous spectrum,
which is $(\pi/2)^2 \approx 2.47$ for $d=2$.
Second, we see that this situation
provides a negative answer to~(Q5),
\ie\ the $\PT$-symmetry can be broken,
and a partial answer to~(Q7).
Let us comment on the behaviour depicted by the cyan curve.
There is a critical value of the parameter~$\eps$ for which
there emerges a pair of complex conjugate eigenvalues
from the continuous spectrum
(we suspect that they emerge due to a collision
of two embedded eigenvalues).
As~$\eps$ increases, the eigenvalues propagate in the complex plane
(this is indicated by the curves joining the blue and red dots
in the right figure;
the dotted curve in the left picture
traces the common real parts of the eigencurves)
till they collide on the real axis and become real.
Then they move on the real axis (as the green dots)
in opposite directions till they reach turning points
(each of them for different value of~$\eps$),
starts to approach each other, coalesce again
and continue as a pair of complex conjugate eigenvalues
(indicated by the black and magenta dots)
until they disappear in the continuous spectrum.
The behaviour of the eigenvalues depicted by the red curve
in the left picture is more difficult in that one of them
seems to have the turning point inside the continuous spectrum.
Because of the collisions we see that the eigenvalues
can actually be degenerate if the $\PT$-symmetry is broken,
providing a positive answer to one of the questions from~(Q6).

Let us mention that the behaviour of the eigenvalues
in the regime of broken $\PT$-symmetry exhibits
certain similarities with the over-damped phenomena
as regards the spectrum of the infinitesimal generator of the semigroup
associated with the damped wave equation
\cite{Cox-Zuazua_1994, Freitas_1999, FK1}.
This indicates the unifying framework of Krein spaces
behind these two problems
\cite{Langer-Tretter_2004, Jacob-Trunk-Winklmeier_2007}.

\begin{figure}[h!]
\epsfig{file=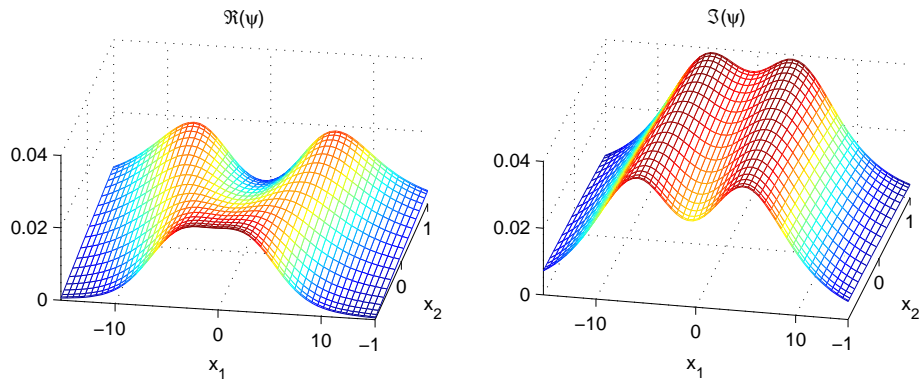, width=\textwidth}
\caption{Real and imaginary parts of an eigenfunctions
corresponding to the smallest (positive) eigenvalue of $H_\alpha$
for $\alpha(x_1)=1/3-0.65 \exp(-0.025 x_1^2)$.}\label{Fig.wavefctn}
\end{figure}

In Figure~\ref{Fig.wavefctn}, we present an example of eigenfunction
corresponding to the case (B1) of Theorem~\ref{Thm2}.
We check that the behaviour of the eigenfunction
is in perfect agreement with the asymptotic results
of Theorem~\ref{Thm3}.

Even if the corresponding eigenenergies are real,
the non-Hermiticity prevents from choosing the eigenfunctions real.
Since the latter is in particular true for the lowest eigenvalue,
it does not make sense to speak about the super- and sub-harmonic
properties of the corresponding eigenfunction
(which hold in the self-adjoint case).
However, although there is no variational characterization of
eigenvalues in the present model, numerically we observe that the
real and imaginary parts of the eigenfunction
corresponding to the lowest eigenvalue
are super-harmonic separately in the regime of unbroken $\PT$-symmetry.
This follows, of course, from the observations
that they do not change sign and that the spectrum is positive.

\begin{figure}[h!]
\epsfig{file=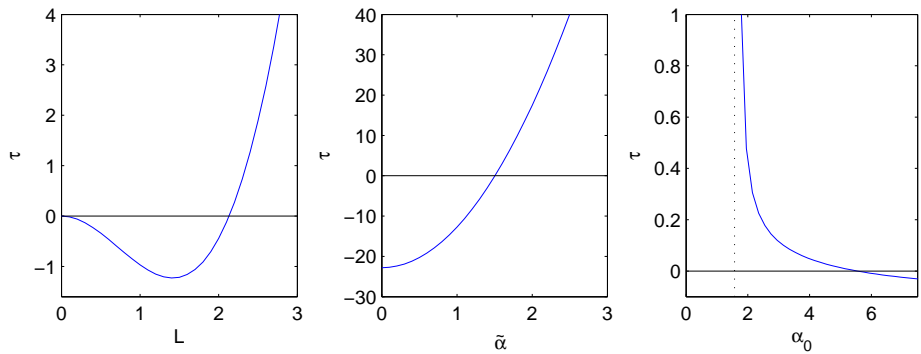, width=\textwidth}
\caption{Dependence of~$\tau$ on parameters defining~$\alpha$
in the step-like situation of Section~\ref{Sec.matching},
with $d=2$.
The first two figures deal with an antisymmetric square well
in the regime $|\alpha_0|<\pi/d$,
while the last one deals with a symmetric square well
in the regime $\alpha_0>\pi/d$.
The first (respectively second) figure
shows the dependence of~$\tau$
on the width $\pm L_\pm=:L$
(respectively on the coupling $\tilde{\alpha}$)
for fixed $\alpha_0=1/3$ and $\alpha_\pm=\alpha_0 \mp 1$
(respectively fixed $\alpha_0=1$, $\pm L_\pm=2$
and variable $\alpha_\pm=\alpha_0\mp\tilde{\alpha}$).
In the last figure we show the dependence of~$\tau$ on~$\alpha_0$
for $L=10$ and variable $\alpha_\pm=(\alpha_0-1)$;
here the dotted line corresponds to $\pi/d$.}\label{Fig.taunegative}
\end{figure}

Finally, in Figure~\ref{Fig.taunegative} we visualize the dependence
of the complicated quantity~$\tau$ defined in~\eqref{tau}
on various parameters. In particular, we see that it changes sign,
giving a positive answer to~(Q1).
Consequently, all the cases of Theorem~\ref{Thm2}
for the critical case $\langle \beta \rangle=0$
and for the regime $|\alpha_0| > \pi/d$ can be achieved.
We also see that the first figure is in qualitative agreement
with an analytic result of~\cite[Prop.~2.1]{BK}.

\section{Conclusion}\label{Sec.end}
%
In this paper we tried to enlarge our knowledge of the point spectrum
of a non-Hermitian $\PT$-symmetric operator introduced in~\cite{BK}
by analyzing it numerically.
We confirmed theoretical results obtained in~\cite{BK}
by perturbation methods,
and showed that they actually hold under much milder conditions about~$\alpha$.

Besides this, it turned out that the operator
can model a fairly wide range of situations.
Indeed, its spectrum is very rich,
and certain properties we found are unusual
when we compare them with the standard self-adjoint cases.
Among them we would like to point out the non-monotonic dependence
on the strength of perturbation
and the existence of the regime of broken $\PT$-symmetry.
We hope that this study will stimulate further theoretical
endeavour to extract and prove the salient features.

In particular, based on the present numerical analysis,
we conjecture that there will be no other spectrum
except for the continuous one if the parameter~$\eps$
is sufficiently large.
At the same time, we were not able to find any discrete eigenvalues
in the case of Corollary~\ref{Corol.reality},
\ie~(Q3) from the end of Section~\ref{Sec.model}
seems to have a negative answer;
the statement~(v) of Theorem~\ref{Thm1} would be trivial, then.
Our numerical experiments also indicate
that the condition mentioning in~(Q2)
is indeed just a technical hypothesis in Theorem~\ref{Thm2}.C,
in the sense that it does not influence the existence/non-existence
of weakly-coupled eigenvalues.

More generally, the existence of eigenvalues in the present model
seems to have a nice heuristic explanation.
We observe that the discrete spectrum behaves in many respects
as that of a one-dimensional Schr\"odinger operator
governed by the first-transverse-eigenvalue potential, \ie,
$
  -\Delta + \min\{\alpha^2,\pi^2/d^2\}
$
in $\sii(\Real)$.
Of course, this self-adjoint idealization is just approximative
and cannot explain, in particular, the existence of non-real eigenvalues.
However, it provides an insight into the non-monotonicity behaviour,
the absence of point spectrum for large~$\eps$,
the positivity of (the real part of) the spectrum, \etc.
It also formally explains the condition
from the statement~1 (respectively~2) of Theorem~\ref{Thm2}.B,
since this actually implies that the potential
is attractive (respectively repulsive).

In this paper we were mainly interested in the eigenvalues
emerging from the threshold~$\mu_0^2$ of the continuous spectrum.
A complete answer to the first question of~(Q7)
can be provided by a perturbation method similar to that of~\cite{BK}.
However, a more detailed analysis of the continuous spectrum
would be still desirable.
In particular, Figure~\ref{Fig.crossing} suggests
that there can be embedded eigenvalues for larger values
of the coupling parameter.

Finally, let us point out that the question of a direct
physical motivation for the Hamiltonian~$H_\alpha$ remains open.
In this paper we were rather interested
in consequences of the non-self-adjointness
on spectral properties of this specific model
in the context of $\PT$-symmetric quantum mechanics.
On the other hand, motivated by problems in semiconductor physics,
similar self-adjoint, respectively non-self-adjoint but dissipative,
Robin-type boundary conditions has been considered recently
in~\cite{Jilek}, respectively in~\cite{KNR}.
In a different context,
the present ($\PT$-symmetric)
Robin-type boundary conditions imply that
we actually deal with the Helmholtz equation in
an electromagnetic waveguide with
radiation/dissipative boundary conditions.

\subsection*{Acknowledgment}
The authors are grateful to Denis Borisov for valuable discussions.
The work has been supported by
the Czech Academy of Sciences and its Grant Agency
within the projects IRP AV0Z10480505 and A100480501,
and by the project LC06002 of the Ministry of Education,
Youth and Sports of the Czech Republic.

%
{\small

\providecommand{\bysame}{\leavevmode\hbox to3em{\hrulefill}\thinspace}
\providecommand{\MR}{\relax\ifhmode\unskip\space\fi MR }
\providecommand{\MRhref}[2]{%
  \href{http://www.ams.org/mathscinet-getitem?mr=#1}{#2}
}
\providecommand{\href}[2]{#2}

}
\end{document}